\documentclass[a4paper]{jpconf}
\usepackage{graphicx}
\begin{document}
\title{Transmuted spectrum-generating algebras and detectable parastatistics of the Superconformal Quantum Mechanics}

\author{Francesco Toppan}

\address{
CBPF, Rua Dr. Xavier Sigaud 150, Urca,
cep 22290-180, Rio de Janeiro (RJ), Brazil.
}

\ead{toppan@cbpf.br}

\begin{abstract}
In a recent paper (Balbino-de Freitas-Rana-FT, arXiv:2309.00965) we proved that the supercharges of the supersymmetric quantum mechanics can be statistically transmuted and accommodated into a $Z_2^n$-graded parastatistics.
In this talk I derive the $6=1+2+3$ transmuted spectrum-generating algebras  (whose respective $Z_2^n$ gradings are $n=0,1,2$) of the ${\cal N}=2$ Superconformal Quantum Mechanics. These spectrum-generating algebras allow to compute, in the corresponding multiparticle sectors of the de Alfaro-Fubini-Furlan deformed oscillator, the degeneracies of each energy level.  The levels induced by the $Z_2\times Z_2$-graded paraparticles cannot be reproduced by the ordinary bosons/fermions statistics. This implies the theoretical detectability of the $Z_2\times Z_2$-graded parastatistics.
\end{abstract}

\section{Introduction}

In recent years the $Z_2^n$-graded, {\it color} Lie algebras and superalgebras introduced in 1978 by Rittenberg-Wyler \cite{{riwy1},{riwy2}} (see also the 1979 paper by Scheunert \cite{sch}) received a boost of attention by both physicists and mathematicians due to several different developments. In particular it was shown that color superalgebras appear as dynamical symmetries of known physical systems as the L\'evy-Leblond spinors \cite{{aktt1},{aktt2}}, while a systematic construction of $Z_2^2$-graded invariant classical \cite{{akt1},{brusigma}} and quantum \cite{{brdu},{akt2},{kuto}} models started (more information and references on recent developments are presented in \cite{aikt} and \cite{nbits}).  For many years progress in this field was hampered by a misconception.
Since $Z_2^n$-graded color Lie (super)algebras can be reconstructed via Klein's operators (for a recent account see e.g. \cite{que}), they were dismissed as not having a direct physical relevance (the same argument, applied to ordinary fermions which, in lower dimensions, can be obtained via bosonization, would imply that fermions are not physically relevant either!).  The connection of $Z_2^n$-graded Lie (super)algebras with a certain special type of parastatistics had been investigated in several works \cite{{yaji},{tol1},{stvj1},{stvj2}}. Till recently, on the other hand,
the question of whether these parastatistics imply {\it inequivocal} different results not reproducible by the ordinary statistics involving bosons and fermions was left unanswered. This question became urgent when the first quantum model invariant under a $Z_2^2$-graded worldline superPoincar\'e algebra was presented by Bruce and Duplij in
\cite{brdu}.  Since the Hamiltonian of that model is also an example of an ordinary ${\cal N}=2$ supersymmetric quantum mechanics, the physical relevance of the $Z_2^2$-graded parastatistics was unclear. A positive answer to this question was finally produced in \cite{top1} (for theories involving $Z_2^2$-graded parafermions) and \cite{top2} 
(for theories involving $Z_2^2$-graded parabosons). It was shown, in a controlled setup, that the eigenvalues of certain observables acting in the multiparticle sectors of such theories allow to determine whether the system under consideration is composed by ordinary particles or by $Z_2^2$-graded paraparticles. The \cite{{top1},{top2}} works employed the Majid's framework \cite{maj} which encodes parastatistics within a graded Hopf algebra endowed with a braided tensor product (the traditional approach to parastatistics makes use of the Green's trilinear relations \cite{gre}; the connection between the two approaches has been discussed in \cite{{anpo},{kada}}). The theoretical detectability of $Z_2^2$-graded parastatistics becomes particularly interesting in the light of the recent experimentalists'advances in either simulating \cite{parasim} or engineering in the laboratory \cite{paraexp} certain types of parastatistics.\par
The \cite{nbits} paper presents several results concerning the classification of $Z_2^n$-graded Lie (super)algebras and associated parastatistics, their construction in terms of Boolean logic gates, invariant hamiltonians under $Z_2^n$-graded (super)algebras, the statistical transmutations of the supercharges of the supersymmetric quantum mechanics. It was further shown that $Z_2^2$-graded paraparticles directly affect (contrary to the previous models discussed in \cite{{top1},{top2}}) the energy spectrum of the superconformal quantum mechanics.  In this talk I discuss a side line which was not addressed in \cite{nbits}, namely the derivation of the energy spectra of the $2$-particle statistical transmutations of the ${\cal N}=2$ de Alfaro-Fubini-Furlan deformed oscillator \cite{dff} as induced by the corresponding transmuted spectrum-generating algebras. There are six different cases: the two pairs
of ${\cal N}=2$ creation/annihilation operators can be assumed to be $2$ fermions ($2F$), $1$ fermion and $1$ boson ($1F+1B$), $2$ bosons ($2B$), or $Z_2^2$ paraparticles, respectively given by $2$ parafermions ($2P_F$),
$1$ parafermion and $1$ paraboson ($1P_F+1P_B$), $2$ parabosons ($2P_B$).  The (para)fermions satisfy the Pauli exclusion principle. The original (not transmuted) spectrum-generating superconformal algebra corresponds to the $2F$ case and is given by $sl(2|1)$ (references for the not-transmuted de Alfaro-Fubini-Furlan ${\cal N}=2$ superconformal quantum mechanics are \cite{{fil},{cht},{ackt}}.

\section{The ${\cal N}=2$ Superconformal Quantum Mechanical model}

In terms of the $2\times 2$ matrices
{\footnotesize{$
I=\left(\begin{array}{cc}1&0\\0&1\end{array}\right),~X=\left(\begin{array}{cc}1&0\\0&-1\end{array}\right),~
Y=\left(\begin{array}{cc}0&1\\1&0\end{array}\right),~ A=\left(\begin{array}{cc}0&1\\-1&0\end{array}\right)
$}}, the ${\cal N}=2$ differential matrix representation of $sl(2|1)$ is given by{\footnotesize{
\begin{eqnarray}
Q_1&=& \frac{1}{\sqrt{2}}\left(\partial_x\cdot A\otimes I +\frac{\beta}{x}\cdot Y\otimes I\right),\nonumber\\
Q_2&=& \frac{1}{\sqrt{2}}\left(\partial_x\cdot Y\otimes A +\frac{\beta}{x}\cdot A\otimes A\right),\nonumber\\
{\widetilde Q}_1&=& \frac{i}{\sqrt{2}} x\cdot A\otimes I,\nonumber\\
{\widetilde Q}_2&=& \frac{i}{\sqrt{2}} x\cdot Y\otimes  A,\nonumber\\
H&=& \frac{1}{2}\left( -\partial_x^2+\frac{\beta^2}{x^2}\right)\cdot I\otimes I -\frac{\beta}{2x^2}\cdot X\otimes I,\nonumber\\
D&=& -\frac{i}{2}\left( x\partial_x+\frac{1}{2}\right)\cdot I\otimes I,\nonumber\\
K&=& \frac{1}{2}x^2\cdot I\otimes I,\nonumber\\
R&=&\frac{i}{4}\left(X\otimes A +2\beta\cdot I\otimes A\right),
\end{eqnarray}
}}
where $R$ is the $R$-symmetry generator. The operators are Hermitian and $\beta $ is an arbitrary real parameter. 
The de Alfaro-Fubini-Furlan \cite{dff} Hamiltonian $H_{DFF}$, introduced through the position
\begin{eqnarray}
H_{DFF} &=& H+K,
\end{eqnarray}
corresponds to a $\beta$-deformation of a matrix quantum oscillator.\par
The $j=1,2$ pairs of creation/annihilation operators $a_j^\dagger, a_j$, defined through
\begin{eqnarray}
a_j := Q_j-i {\widetilde Q}_j,  &\qquad &  a_j^\dagger := Q_j+i{\widetilde Q}_j,
\end{eqnarray}
satisfy
\begin{eqnarray}\label{hamaadag}
[H_{DFF}, a_j] = -a_j, &\quad & [H_{DFF},a_j^\dagger] = a_j^\dagger.
\end{eqnarray}

\section{The six statistical transmutations}

The two creation operators $a_1^\dagger, a_2^\dagger$ can be assigned to be:\\
- both bosonic $(2B)$, corresponding to an ordinary $Z_2^0$-graded Lie algebra, \\
- one bosonic and one fermionic $(1B+1F)$ or both fermionic $(2F)$ (corresponding to ordinary $Z_2^1$-graded superalgebras or,\\
- expressed by $Z_2^2$-graded parastatistics according to the two tables below given by 
\begin{eqnarray} \label{casea}
{\textrm{the ${Z}_2^2$ color Lie algebra:}}\qquad&& 
\begin{array}{c|cccc}&00&10&01&11\\ \hline 00&0&0&0&0\\10&0&0&1&1\\01&0&1&0&1\\11&0&1&1&0
\end{array}
\end{eqnarray}
\begin{eqnarray}\label{caseb}
{\textrm{the ${Z}_2^2$ color Lie superalgebra:}}\qquad&& 
\begin{array}{c|cccc}&00&10&01&11\\ \hline 00&0&0&0&0\\10&0&1&0&1\\01&0&0&1&1\\11&0&1&1&0
\end{array}\qquad
 {\textrm{}}
\end{eqnarray}
(the $0,1$ entries respectively denote
commutators or anticommutators of the corresponding $2$-bit particles belonging to the sectors denoted as $00,10,01,11$).  Let $A,B$ two operators of respective $2$-bit gradings $\alpha, \beta$. Their (anti)commutator
$(A,B)$ 
is defined as $(A,B)=AB-(-1)^{\varepsilon_{\alpha\beta}}BA$, where $\varepsilon_{\alpha\beta}=0,1$ is given by the corresponding entry in tables (\ref{casea}, \ref{caseb}). The grading of $(A,B)$ is $\alpha+\beta ~mod~2$.\par
The $2P_B$ parastatistics is recovered by assigning, let's say, $a_1^\dagger \in 10$, $a_2^\dagger\in 01$ from (\ref{casea}); the $1P_B+1P_F$ and $2P_F$ parastatistics are recovered from (\ref{caseb}) with respective assignments given by $a_1^\dagger \in 10$, $a_2^\dagger\in 11$ and $a_1^\dagger \in 10$, $a_2^\dagger\in 01$. In these three $Z_2^2$-graded cases the Hamiltonian $H_{DFF}$ is assigned to the (bosonic) $00$-sector.

\section{The multiparticle sectors}
Following \cite{{maj},{top1},{top2}}, the $Z_2^n$-graded parastatistics of a multiparticle sector is encoded in a graded Hopf algebra endowed with a braided tensor product.\par
Let  $A,B,C,D$ be ${Z}_2^n$-graded operators whose respective $n$-bit gradings are $\alpha,\beta,\gamma,\delta$. The braided tensor product, conveniently denoted as ``${\otimes}_{br}$", is defined to satisfy
the relation
\begin{eqnarray}\label{braidedtensor}
(A\otimes_{br} B)\cdot (C\otimes_{br} D) &=& (-1)^{\langle\beta,\gamma\rangle}(AC)\otimes_{br} (BD),
\end{eqnarray}
in terms of a $(-1)^{\langle\beta,\gamma\rangle}$ sign.  For $Z_2^2$, the $0,1$ values of ${\langle\beta,\gamma\rangle}$ are read from the tables (\ref{casea}) and (\ref{caseb}).  The coproduct $\Delta$ is the relevant Hopf algebra operation which allows, in physical applications, to construct multiparticle states. For a Universal Enveloping Algebra $U\equiv {U}({\cal G})$ of a graded Lie algebra ${\cal G}$ the coproduct map, given by
\begin{eqnarray}\label{coproduct}
\Delta &:& U\rightarrow U\otimes_{br} U,
\end{eqnarray} 
satisfies the coassociativity property
\begin{eqnarray}\label{coassoc}
 &\Delta^{m+1}:=  (\Delta\otimes_{br} {\mathbf 1})\Delta^m=({\mathbf 1}\otimes_{br} \Delta)\Delta^m \qquad ~ {\textrm{(where~ $\Delta^1\equiv \Delta$)}}
\end{eqnarray}
and the comultiplication
\begin{eqnarray}\label{comult}
\Delta(u_1u_2) &=& \Delta(u_1)\cdot \Delta(u_2) \qquad {\textrm{for any ~$u_1,u_2\in U$.}}
\end{eqnarray}
The action of the coproduct on the identity ${\bf 1}\in {U}({{\cal G}})$ and on the primitive elements $g\in{{\cal G}}$ is
\begin{eqnarray}\label{coproductaction}
\Delta({\bf 1})={\bf 1}\otimes_{br}{\bf 1}, \quad&&\quad
\Delta(g) = {\bf 1}\otimes_{br} g+g\otimes_{br} {\bf 1}.
\end{eqnarray} 
In physical applications, typical primitive elements are the Hamiltonians and the creation/annihilation operators.\par
For $\beta>-\frac{1}{2}$ the single-particle Hilbert space ${\cal H}_{\beta}^{(1)}$ is spanned by repeatedly applying the $a_1^\dagger, a_2^\dagger$ creation operators on the normalized single-particle Fock vacuum  $\Psi_{\beta}(x)\equiv |vac\rangle_1$:
{\footnotesize{\begin{eqnarray}
\Psi_{\beta}(x) &=& \frac{1}{\sqrt{\Gamma(\beta+\frac{1}{2})}}x^\beta e^{-\frac{1}{2}x^2}\left(\begin{array}{c}1\\0\\0\\0\end{array}\right), \qquad {\textrm{with}}\quad a_1\Psi_\beta(x)=a_2\Psi_\beta(x)=0.
\end{eqnarray}}}
Similarly, the  $2$-particle Hilbert space ${\cal H}_{\beta}^{(2)}$ is spanned by repeatedly applying the $\Delta(a_1^\dagger),\Delta( a_2^\dagger)$ creation operators on the normalized $2$-particle Fock vacuum $\Psi_{\beta}(x,y)\equiv |vac\rangle_2$:
{\footnotesize{\begin{eqnarray}
\Psi_{\beta}(x,y)&=&\frac{1}{\Gamma(\beta+\frac{1}{2})}(xy)^\beta e^{-\frac{1}{2}(x^2+y^2)}\rho_1,\qquad {\textrm{with}}\quad\Delta( a_1)\Psi_\beta(x,y)=\Delta(a_2)\Psi_\beta(x,y)=0,
\end{eqnarray}}}
where $\rho_1$ is the $16$-component vector with entry $1$ in the first position and $0$ otherwise.\par
The single-particle and $2$-particle vacuum energy are respectively $\frac{1}{2}+\beta$ and $1+2\beta$:
\begin{eqnarray}
H_{DFF} |vac\rangle_1 = (\frac{1}{2}+\beta)|vac\rangle_1, && 
\Delta(H_{DFF}) |vac\rangle_2 = (1+2\beta)|vac\rangle_2.
\end{eqnarray}

\section{The six transmuted spectrum-generating graded(super)algebras} 

Let us set for convenience $P=a_1^\dagger$, $Q=a_2^\dagger$ (in the single-particle sector) and ${\overline P}=\Delta_\ast(a_1^\dagger)$, ${\overline Q}=\Delta_\ast(a_2^\dagger)$ (in the $2$-particle sector). The asterisk in the coproduct denotes one of the six (para)statistics $2B, 1B+1F, 2F, 2P_B, 1P_B+1P_F, 2P_F$ defined by the respective signs entering (\ref{braidedtensor}). The construction of ${\overline P}, {\overline Q}$ is not affected by the signs. The signs specify how ${\overline P}, {\overline Q}$ are interchanged. The coproduct guarantees the homomorphism of the graded Lie algebras
(defined in terms of (anti)commutators) of the single-particle and $2$-particle sectors. We present the six transmuted spectrum-generating algebras in terms of the ${\overline P}, {\overline Q}$ $2$-particle generators. 
In the $2B$ and $1B+1F$ cases they close as non-linear (super)algebras satisfying (anti)symmetry and graded Jacobi identities. In the remaining cases the (super)algebras close linearly on a finite number of generators. The spectrum-generating (super)algebras are defined by the following sets of (anti)commutation relations:
\begin{eqnarray}\label{twob}
2B:  \qquad\qquad &&\nonumber\\
\relax[{\overline P}, {\overline Q}] &= &{\overline W},\nonumber\\
\relax [{\overline P}, {\overline W}] &=& (1-z){\overline Q}^3 +z {\overline P}^2{\overline Q}-(3-z){\overline P}{\overline Q}{\overline P} +z{\overline Q}{\overline P}^2,\nonumber\\
\relax [{\overline Q}, {\overline W}] &=&- (1+u){\overline P}^3 +u {\overline Q}^2{\overline P}+(3+u){\overline Q}{\overline P}{\overline Q} +u{\overline P}{\overline Q}^2,\nonumber\\
\end{eqnarray}
for arbitrary $u,z$ values. A convenient choice is to set $u=z=0$.
\begin{eqnarray}
1B+1F:  \qquad\qquad &&\nonumber\\
\relax\{{\overline P}, {\overline P}\} &= &{\overline Z},\nonumber\\
\relax[{\overline P}, {\overline Q}] &= &{\overline W},\nonumber\\
\relax[{\overline W}, {\overline W}] &= &{\overline C},\nonumber\\
\relax\{{\overline P}, {\overline W}\} &= &{\overline 0},\nonumber\\
\relax [{\overline Q}, {\overline W}] &=& -(2+u){\overline P}^3 +u {\overline P}{\overline Q}^2+2{\overline Q}{\overline P}{\overline Q} +u{\overline Q}^2{\overline P},\nonumber\\
\relax [{\overline Z}, \ast] &=&0,\nonumber\\
\relax [{\overline C}, \ast] &=&0,
\end{eqnarray}
for arbitrary values of $u$; ${\overline P}$ is a fermionic generator, while ${\overline Q}$ is bosonic.
\begin{eqnarray}
2F:  \qquad\qquad &&\nonumber\\
\relax\{{\overline P}, {\overline P}\} &= &\{{\overline Q}, {\overline Q}\}= {\overline Z},\nonumber\\
\relax [{\overline Z}, {\overline P}] &=& [{\overline Z}, {\overline Q}]~~ =0;
\end{eqnarray}
it is the original superalgebra of the ${\cal N}=2$ superconformal creation operators.\par
The three $Z_2^2$-graded superalgebras are 
\begin{eqnarray}
2P_B:  \qquad\qquad &&\nonumber\\
\relax\{{\overline P}, {\overline Q}\} &= &0;
\end{eqnarray}
since ${\overline P}, {\overline Q}$ are parabosonic, it is defined by  a single anticommutator.
\begin{eqnarray}
1P_B+ 1P_F:  \qquad\qquad &&\nonumber\\
\relax\{{\overline P}, {\overline P}\} &= &{\overline Z},\nonumber\\
\{{\overline P},{\overline Q}\} &=& 0,\nonumber\\
\relax [{\overline Z}, \ast] &=&0,
\end{eqnarray}
where ${\overline P}$ is the parafermionic operator.
\begin{eqnarray}
2P_F:  \qquad\qquad &&\nonumber\\
\relax\{{\overline P}, {\overline P}\} &= &\{{\overline Q},{\overline Q}\}={\overline Z},\nonumber\\
\relax [{\overline P},{\overline Q}] &=& {\overline W},\nonumber\\
\{{\overline P},{\overline W}\} &=& \{{\overline Q},{\overline W}\} =0,\nonumber\\
\relax [{\overline Z}, \ast] &=&0.
\end{eqnarray}
In all above $6$ cases it follows from (\ref{hamaadag}) that, by setting ${\overline H}_{DFF} =\Delta (H_{DFF})$, ${\overline P}, {\overline Q}$ create a quantum of energy:
\begin{eqnarray}
\relax [{\overline H}_{DFF}, {\overline P}]={\overline P}, && [{\overline H}_{DFF}, {\overline Q}]={\overline Q}.
\end{eqnarray}
We now compute the degeneracies of the $2$-particle energy levels implied by the above $6$ graded (super)algebras.

\section{Degeneracies of the $2$-particle energy levels}

In the $2B$ case the set of linearly independent operators that create the $n_1+n_2+n_3$ excited states from the $2$-particle  vacuum $|vac\rangle_2$ is given by the ordered operators
\begin{eqnarray}\label{ordered}
&{\overline P}^{n_1}{\overline W}^{n_2}{\overline Q}^{n_3},&
\end{eqnarray}
where $n_1,n_3$ are non-negative integers, while $n_2$ is restricted to be $n_2=0,1$.\par
Indeed, the first relation in (\ref{twob}) states that ${\overline P}{\overline Q}-{\overline Q}{\overline P}={\overline W}$, so that ${\overline Q}$ can be put at the right of ${\overline P}$:
\begin{eqnarray}\label{sim1}
{\overline Q}{\overline P} &\sim& {\overline P}{\overline Q}, {\overline W}.
\end{eqnarray}
Furthermore, by setting $z=u=0$, ${\overline P}{\overline W}-{\overline W}{\overline P} = {\overline Q}^3-3{\overline P}{\overline Q}{\overline P}$, combined with (\ref{sim1}), implies that
 \begin{eqnarray}\label{sim2}
{\overline W}{\overline P} &\sim& {\overline P}{\overline W}, {\overline Q}^3, {\overline P}^2{\overline Q},
\end{eqnarray}
so that ${\overline W}$ can be put on the right of ${\overline P}$.
Next, ${\overline Q}{\overline W}-{\overline W}{\overline Q} =-{\overline P}^3+3{\overline Q}{\overline P}{\overline Q}$ implies, together with (\ref{sim1}), that
\begin{eqnarray}{\label{sim3}}
{\overline Q}{\overline W}&\sim& {\overline W}{\overline Q}, {\overline P}^3, {\overline P}{\overline Q}^2,
\end{eqnarray}
so that ${\overline Q}$ can be put on the right of ${\overline W}$.\par
Finally, the $n_2=0,1$ restriction comes from the ${\overline W}^2 =({\overline P}{\overline Q}-{\overline Q}{\overline P})({\overline P}{\overline Q}-{\overline Q}{\overline P})$ relation.\par
The ordered expression (\ref{ordered}) easily allows to evaluate the degeneracy $n_{deg}(L)$ of the $L=n_1+n_2+n_3$ excited energy states ${\overline P}^{n_1}{\overline W}^{n_2}{\overline Q}^{n_3}|vac\rangle_2$. For $L>0$ it is given by $n_{deg}(L)=2L$.\par
In the $2P_B$ case, due to the single relation ${\overline P}{\overline Q}= -{\overline Q}{\overline P}$, the linearly independent operators which create the excited energy states of level $L$ are ${\overline P}^n{\overline Q}^{L-n}$ for $n=0,1,\ldots, L$. It follows that its $n_{deg}(L)$ degeneracy is $L=n_1+n_2$. The similar analysis can be produced for all other $4$ cases. In the original, not-transmuted, $2F$ case, due to the ${\cal N}=2$ supersymmetry the degeneracy $n_{deg}(L)$ of the excited states of energy level $L>0$ is $n_{deg}(L)=2$. In all $6$ cases the energy spectrum of the $2$-particle sector is given by $1+2\beta+L$, for $L\geq 0$. The vacuum at $L=0$ is not degenerate ($n_{deg}(L=0) =1$). The $L>0$ degeneracies are given in the table
\begin{eqnarray}
2B \quad {\textrm{and}} \quad 1B+1F&:& n_{deg}(L) = 2L,\nonumber\\
2F \qquad\qquad&:& n_{deg}(L) = 2,\nonumber\\
2P_B, ~1P_B+1P_F, ~2P_F &:& n_{deg}(L) = L+1.
\end{eqnarray}
Even if the analysis of the $2$-particle energy spectrum of the theory does not allow to discriminate the three $Z_2^2$-graded parastatistics $2P_B, 1P_B+1P_F, 2P_F$,  it allows to single out them with respect to the ordinary bosons/fermions statistics. This is sufficient to prove the theoretical detectability of the $Z_2^2$-graded paraparticles in the framework of the statistically transmuted superconformal quantum mechanics.\par
The investigations about the detectability of the $Z_2^2$-graded  parastatistics (and, more generally, $Z_2^n$-graded paraparticles described by $n$ bits), even if just at the beginning are promising. The aim is to find some realistic model which could be put into experimental test.


\section*{References}
\begin{thebibliography}{9}
\bibitem{riwy1} Rittenberg V and Wyler D 1978, Generalized Superalgebras, 
{\it Nucl. Phys.} B {\bf 139}, 189.
\bibitem{riwy2} Rittenberg V and Wyler D 1978, Sequences of $Z_2\otimes Z_2$ graded Lie algebras and superalgebras
{\it J. Math. Phys.} {\bf 19}, 2193.
\bibitem{sch} Scheunert M 1979, 
{Generalized Lie algebras}
{\it J. Math. Phys.} {\bf 20}, 712.
\bibitem{aktt1} Aizawa N, Kuznetsova Z, Tanaka H and Toppan F 2016, {$ {Z}_2 \times {Z}_2$-graded Lie symmetries of the L\'evy-Leblond equations}, {\it Prog. Theor. Exp. Phys.} {\bf 123A01} ({\it Preprint} arXiv:1609.08224).
\bibitem{aktt2} Aizawa N, Kuznetsova Z, Tanaka H and Toppan F 2017, {Generalized supersymmetry and L\'evy-Leblond equation}, in  {\it Physical and Mathematical Aspects of Symmetries} eds S. Duarte et al (Springer, Cham), p. 79 ({\it Preprint} arXiv:1609.08760).
\bibitem{akt1} Aizawa N, Kuznetsova Z and Toppan F 2020, {${Z}_2\times{Z}_2$-graded mechanics: the classical theory} {\it Eur. J. Phys.} C {\bf 80}, 668 ({\it Preprint} arXiv:2003.06470).
\bibitem{brusigma} Bruce A J 2020, {${Z}_2\times{Z}_2$-graded supersymmetry: 2-d sigma models} {{\it J. Phys. A: Math. Theor.}} {\bf 53}, 455201 ({\it Preprint} arXiv:2006.08169).
\bibitem{brdu} Bruce A J and Duplij S 2020, Double-graded supersymmetric quantum mechanics {\it J. Math. Phys.} {\bf 61}, 063503 ({\it Preprint} arXiv:1904.06975).
\bibitem{akt2} Aizawa N, Kuznetsova Z and Toppan F 2021, {${Z}_2\times {Z}_2$-graded mechanics: the quantization} {\it Nucl. Phys.} B {\bf 967}, 115426 ({\it Preprint} arXiv:2005.10759).
\bibitem{kuto} Kuznetsova Z and Toppan F 2021, {
Classification of minimal $Z_2\times Z_2$-graded Lie (super)algebras and some applications} {\it 
J. Math. Phys.} {\bf 62}, 063512 ({\it Preprint} arXiv:2103.04385).
\bibitem{aikt} Aizawa N, Ito R, Kuznetsova Z and Toppan F 2023, {New aspects of the ${Z}_2\times{Z}_2$-graded $1D$ superspace: closed strings and $2D$ relativistic models} {\it Nucl. Phys.} B {\bf 991}, 116202 ({\it Preprint} arXiv:2301.06089).
\bibitem{nbits} Balbino M M, de Freitas I P, Rana R G and Toppan F 2023, Inequivalent ${Z}_2^n$-graded brackets, $n$-bit parastatistics and statistical transmutations of supersymmetric quantum mechanics {\it Preprint}
arXiv:2309.00965.
\bibitem{que} Quesne C 2021, {Minimal bosonization of double-graded supersymmetric quantum mechanics} {\it Mod. Phys. Lett.} A {\bf 36}, 2150238 ({\it Preprint} arXiv:2108.06243). 
\bibitem{yaji} Yang W M and Jing S C 2001, {A new kind of graded Lie algebra and parastatistical supersymmetry}
{\it Sci. in China (Series A)} {\bf 44}, 9 ({\it Preprint} arXiv:math-ph/0212004). 
\bibitem{tol1} Tolstoy V N 2014, {Once more on parastatistics} {\it Phys. Part. Nucl. Lett.} {\bf{11}},  933 ({\it Preprint} arXiv:1610.01628[math-ph]).
\bibitem{stvj1} Stoilova N I and Van der Jeugt J 2018, {The ${Z}_2\times{Z}_2$-graded Lie superalgebra $pso(2m+1|2n)$ and new parastatistics representations} {\it  J. Phys. {A}: Math. Theor.} {\bf 51}, 135201 ({\it Preprint} arXiv:1711.02136).
\bibitem{stvj2} Stoilova N I and Van der Jeugt J 2022, {The ${{Z}}_{2}\times {{Z}}_{2}$-graded Lie superalgebras ${p}{s}{o}(2n+1\vert 2n)$ and ${p}{s}{o}(\infty \vert \infty )$, and parastatistics Fock spaces}  {\it J. Phys. A: Math. Theor.} \textbf{55}, 045201 ({\it Preprint} arXiv:2112.12811). 
\bibitem{top1} Toppan F 2021, {$Z_2\times Z_2$-graded parastatistics in multiparticle quantum Hamiltonians} {\it J. Phys. A: Math. Theor.} {\bf 54}, 115203 ({\it Preprint} arXiv:2008.11554).
\bibitem{top2} Toppan F 2021, {Inequivalent quantizations from gradings and $Z_2\times Z_2$-graded parabosons} {\it J. Phys. A: Math. Theor.}  {\bf 54}, 355202 ({\it Preprint} arXiv:2104.09692).
\bibitem{maj} Majid S 1995 {\it Foundations of Quantum Group Theory} (Cambridge: Cambridge University Press).
\bibitem{gre} Green H S 1953, A Generalized Method of Field Quantization {\it Phys. Rev.} {\bf 90}, 270.
\bibitem{anpo} Aneva B and Popov T 2005, Hopf Structure and Green Ansatz of Deformed Parastatistics Algebras
{\it J. Phys. A: Math. Gen.} {\bf 38}, 6473 ({\it Preprint} arXiv:math-ph/0412016).
\bibitem{kada} Kanakoglou K and Daskaloyannis C 2007, Parabosons quotients. A braided look at Green's ansatz and a generalization {\it J. Math. Phys.} {\bf  48}, 113516 ({\it Preprint} arXiv:0901.04320).
\bibitem{parasim} Huerta Alderete C and Rodr\'\i guez-Lara B M 2017, Quantum simulation of driven para-Bose oscillators {\it Phys. Rev.} A {\bf 95}, 013820 ({\it Preprint} arXiv:1609.09166).
\bibitem{paraexp} Huerta Alderete C, Greene A M, Nguyen N H, Zhu Y, Rodr\'\i guez-Lara B M and Linke N M 2021,
 Experimental realization of para-particle oscillators {\it Preprint} arXiv:2108.05471.
\bibitem{dff} de Alfaro V, Fubini S and Furlan G 1976, {Conformal invariance in quantum mechanics} {\it Nuovo Cim.} A {\bf 34}, 569.
\bibitem{fil} Fedoruk S, Ivanov E and Lechtenfeld O 2012, {Superconformal mechanics} {\it J. Phys. A: Math. Theor.} {\bf 45}, 173001 ({\it Preprint} arXiv:1112.1947).
\bibitem{cht} Cunha I E, Holanda N L and Toppan F 2017, {From worldline to quantum superconformal mechanics with and without oscillatorial terms: $D(2,1;\alpha)$ and $sl(2|1)$ models}  {\it Phys. Rev.} D {\bf{96}}, 065014 ({\it Preprint} arXiv:1610.07205).
\bibitem{ackt} Aizawa N, Cunha I E, Kuznetsova Z and Toppan F 2019, {On the spectrum-generating superalgebras of the deformed 
one-dimensional quantum oscillators} {\it J. Math. Phys.} {\bf 60}, 042102 ({\it Preprint} arXiv:1812.00873).
\end{thebibliography}
\end{document}